\documentstyle[prc,psfig,aps,twocolumn]{revtex}
\begin{document}
\twocolumn[
\hsize\textwidth\columnwidth\hsize\csname @twocolumnfalse\endcsname

\draft
\begin{center}
PHYSICAL REVIEW C (in press)
\end{center}
\title{Nucleation versus Spinodal decomposition in a first order quark hadron
phase transition}
\author{P. Shukla and A. K. Mohanty}
\address{ Nuclear Physics Division,
Bhabha Atomic Research Centre,\\
Trombay, Mumbai 400 085, India}
\maketitle
\begin{center}
(Received 10 May 2001)
\end{center}
\begin{abstract}
\noindent
 We investigate the scenario of homogeneous nucleation for a first order
quark-hadron phase transition in a rapidly expanding background of
quark gluon plasma. Using an improved preexponential factor
for homogeneous nucleation rate, we solve a set of coupled equations to study
the hadronization and the hydrodynamical evolution of the matter.
It is found that significant
supercooling is possible before hadronization begins. This study also
suggests that spinodal decomposition competes with nucleation and may provide
an alternative mechanism for phase conversion particularly if the
transition is strong enough and the medium is nonviscous. For weak
enough transition, the phase conversion may still proceed via homogeneous
nucleation.

PACS number(s): 12.38.Mh, 64.60.Qb
\end{abstract}
]

\narrowtext

\section{ Introduction }

The hadronization of Quark Gluon Plasma (QGP) possibly
produced in the early universe or expected to be formed
in relativistic heavy-ion collisions \cite{HARRIS} has been the focus of
much attention during the past few years.
The quark gluon plasma (QGP), if formed, would expand hydrodynamically
and would cool down until it reaches a critical temperature  $T_C$ where
a phase transition from the quark matter to the hadron matter begins.
Although the plasma has to hadronize, the mechanism of hadronization
still remains an open question. The percolation model calculations are
used in the case of a second order phase transition. In a first order
scenario, the dynamics of the phase transition has been modeled in several
ways. In the most idealized picture, the temperature of the plasma is held
fixed at $T=T_C$ until the phase conversion is completely over. Assuming
an isoentropic expansion in (1+1) dimension, the hadronization in the above
picture gets completed at $\tau_h=r\,\tau_C$ where $r$ is the ratio of the
degrees of freedom of QGP and hadronic phases and $\tau_C$ is the proper time
at which the QGP cools down to the temperature $T_C$. In reality, a first
order phase transition is characterized by a large nucleation barrier
that separates the two phases at $T=T_C$ and the hadronization will
not begin unless the matter supercools below $T_C$.
Alternatively, the theory of homogeneous nucleation
has been invoked to study the first order phase transition which is more
realistic than the above idealized adiabatic scenario and
has been in use for quite some time in the
cosmological context \cite{FULLER}.
In this picture, the transition is initiated by the nucleation
of critical-size hadron bubbles from a supercooled metastable QGP phase.
These hadron bubbles can grow against surface tension, converting
the QGP phase into the hadron phase as the temperature drops below the
critical temperature, $T_C$.
For strong enough transition, the large amplitude
fluctuations are suppressed so that the nucleation begins from a (nearly)
homogeneous background of supercooled metastable phase. This has been the
basis of homogeneous nucleation theory \cite{LANG} based on which the QCD
phase transition has been studied extensively
\cite{CSER1,RUGGERI,CSER2,ZABPL,ZABPRC,SHUK}.
However, for a weak enough transition, the matter may not remain in a pure
homogeneous state even at $T=T_C$ due to the transitions that may occur
above $T_C$. Evidence of such pretransitional phenomena are quite
common in several condensed
matter studies particularly in the case of isotropic to nematic transition
in liquid crystals.
The possibility of such a transition had also been investigated
during the cosmological electroweak phase transition in the early universe
\cite{GLE1,GLE2}.
Due to high rate of thermal fluctuations and slow cooling of the universe,
a strong phase mixing is expected even at temperature above $T_C$ where the
new phase is highly metastable. In such situation, the phase transition either
proceeds through percolation \cite{GLE2}, if not, the dynamics of the
first order transition is going to be quite different than the standard
theory of homogeneous
nucleation \cite{GLE3}. Similar phenomena is also expected in the
case of a quark hadron phase transition both in the early universe as well as
in the plasma expected to be formed during the relativistic
heavy ion collisions \cite{AGAR,INHOMO}.
An ideal quark gluon plasma in (1+1) dimension expands as per the
Bjorken scaling
where $T^3\tau$ is constant \cite{BJOR}. This scaling would mean that the rate
of change of temperature is higher for the plasma at earlier time
as compared to the plasma at later time. In the case of early universe,
due to high initial temperature, the rate of cooling becomes quite
slow by the time it approaches $T_C$. However, the QGP produced
during the heavy ion collisions will have smaller initial temperature
and will cool much more rapidly as compared to the early universe.
Another difference as compared to the early universe is that the
QGP produced at RHIC and LHC energies may attain kinetic equilibrium
in a very short time $\approx 0.3-0.7$ fm/$c$ but will remain
far off from chemical equilibrium \cite{SHUR,BIRO}.  Such a plasma will be
more gluon rich and many more quark and anti-quark pairs will be needed
before the plasma achieves chemical equilibrium.
The chemically unsaturated plasma
will cool still at a faster rate since additional energy will be consumed in
approaching the chemical equilibration \cite{BIRO,DUTTA}.
Recently, we have investigated the effect of
thermal fluctuations leading to phase mixing by modeling them as
subcritical hadron bubbles \cite{AKM}.
Although the equilibrium density distribution of these subcritical
bubbles can be quite large, their equilibration time scale is larger than
the cooling time scale for the QGP. As a consequence, for RHIC and LHC
energies, they will not build up to a level capable of modifying the
predictions from homogeneous nucleation theory:
the QGP has to supercool and the hadronization may still proceed through
the nucleation of critical bubbles within a nearly homogeneous
background of quark-gluon plasma.

For the QGP formed at RHIC and LHC energies, the question to be asked
is how long the plasma would continue to supercool.
It is expected that the plasma will
supercool upto some temperature $T_m$ until the density of nucleated hadron
bubbles become sufficient to heat up the medium due to the release of latent
heat \cite{CSER2}. Since the medium is heated up, the nucleation is
again switched off at some point and the hadronization continues
due to the growth of previously nucleated hadron bubbles.
Related to this another crucial aspect which needs investigation is
that how fast the barrier between the metastable QGP phase and the stable
hadron phase decreases as the system supercools.
 For strong enough supercooling,
the barrier between the stable and metastable minima completely vanishes
leading to a point of inflection at $T=T_S$ known as spinodal instability.
Thus, the rapidly quenched system leaves the region of metastability
and enters the highly unstable spinodal region before a
substantial amount of nucleation begins.
 The spinodal decomposition has also been suggested as a
possible mechanism of phase conversion for a rapidly expanding system of quarks
and gluons \cite{DUMHEP}. The second scenario of phase transition may lead
to coherent pion emission due to the formation of disoriented chiral
condensates \cite{DUMPRL}.

For the case of QGP produced in heavy ion collision,
the route to hadronization either through nucleation or
through spinodal decomposition
depends sensitively on several factors like nucleation rate,
effective potential used to
model the dynamics of the phase transition and also on the hydrodynamical
evolution that decides the expansion rate of the matter. The homogeneous
nucleation rate is estimated from the factor $I=I_0~\exp(-\Delta F_C/T)$ where
$I_0$ is the product of dynamical and statistical prefactors and $\Delta F_C$
is the minimum energy needed to create a critical hadron bubble.
The exponential factor dominates the rate at temperature close to
$T_C$ whereas the prefactor
$I_0$ plays the significant role away from $T_C$. The most commonly used
expression for the prefactor is $I_0 \sim T^4$ \cite{LINDE} which may
be alright for the case of early universe but not for the heavy ion
collisions where we expect large amount of supercooling.
The theory of homogeneous nucleation assumed significance for the
study of phase transition in QGP produced in
the Relativistic heavy ion collision after Csernai and Kapusta \cite{CSER1}
derived an improved prefactor using
coarse grained field theory. An important aspect of their formalism is the
presence of viscosity which is essential for the removal of latent heat away
from the hadron bubble. The prefactor vanishes for an ideal plasma.
This would suggest that a dissipative dynamics should
be followed for the phase transition to be completed through
nucleation \cite{SHUK}.
In a subsequent work \cite{RUGGERI}, Ruggeri and Friedman, on the other hand,
derived a prefactor which does not vanish in the limit of zero viscosity.
Recently, we have also derived a prefactor which is more general and also
reproduces the Csernai-Kapusta and Ruggeri-Friedman results in the limiting
cases \cite{DYNA}. The prefactor $I_0$ in all these estimates is found to be
less as compared to $T^4$. In this work, we estimate the nucleation rate
using an improved prefactor as derived in Ref.~\cite{DYNA}.
The argument $\Delta F_C/T$
that appears in the exponential depends on the phenomenological potential
model used to describe a first order phase transition. Since the lattice
predictions \cite{IWA96,IWA94} are not yet conclusive about the nature and
the strength of the transition, we use an effective potential that follows the
bag equation of state and also covers a wide range from very weak to very
strong first order phase transitions. Finally, we study amount of
supercooling and
rate of hadronization by solving self consistently the nucleation rate along
with energy momentum conserving hydrodynamic equation both for dissipative
and nondissipative plasma. The dissipative plasma makes the evolution slow
and also generates extra entropy.
From this study we show that spinodal decomposition may compete with
homogeneous nucleation if the plasma is nondissipative and the transition
is relatively stronger. Since the amount of supercooling
is less in case of a dissipative plasma, the phase conversion may still proceed
through the nucleation of critical size hadron bubbles from a homogeneous
background of supercooled quark gluon plasma.

The paper is organized as follows: In Sec.~II, we briefly review the
bag equation of state and an effective potential used to model the phase
transition.
The nucleation rate with various prefactors are discussed in Sec.~III.
Finally, we present our results in Sec.~IV followed by the
conclusion in Sec.~V.

\section{The effective potential and the equation of state}

The spinodal instability corresponds to the inflection point
in the effective potential (below $T_C$) used to study the dynamics
of a phase transition. Obviously, the temperature $T_S$ at which the
instability occurs will depend on the parameters of the effective potential.
Since the order as well as strength  of the quark hadron phase transition
is still an unsettled issue, we consider a more generic form of the
potential which covers a wide range from very strong to very weak first
order phase transition. The parameters of the effective potential are
determined from the physical observables like the surface tension $\sigma$,
the correlation length $\xi$ and also from the requirement that the potential
difference between the two minima should corresponds to the pressure
difference between the quark and hadron phases. We follow the bag equation of
state to estimate various thermodynamical observables. For example,
the pressure can be estimated from the  relation $P=T \Sigma \ln Z_i/V - B$
where the partition function $\ln Z_i$ for a single fermion or
boson is given by \cite{MAD}
\begin{eqnarray}
\ln Z_i&=& \pm \frac{gV}{2\pi^2}
        \int_{0}^{\infty}\,dk\, k^2 \left[
        \ln\left ( 1 \pm e^{-\sqrt{k^2 + m_i^2}/T} \right )\right],
\end{eqnarray}
where the $+$ and $-$ stand for fermions and bosons respectively.
In the above, the
chemical potential $\mu$ has been set to zero. Assuming, a baryon
free quark gluon plasma which consists of $u$, $d$, and $s$ quarks
and gluons, the total pressure can be written as

\begin{eqnarray}\label{pq}
p_q&=& \frac{16\,\pi^2}{90} T^4 +\nonumber \\
& & \sum_{i=1}^{n_f} \frac{12 \,T}{2\pi^2} \int_{0}^{\infty}\,dk\, k^2
          \ln\left ( 1 + e^{-\sqrt{k^2 + m_i^2}/T} \right ) - B, \nonumber \\
    &=& \frac{16\,\pi^2}{90} T^4 +
    \frac{7 \pi^2}{60} T^4 \sum_{i=1}^{n_f} f(m_i/T) - B,
\end{eqnarray}
where $n_f$ is the number of flavours and $m_i$, ($i$=$u$, $d$, $s$)
are their masses.
The function $f(m_i/T) = (360/7\pi^4) \int_{0}^{\infty}\,du\, u^2
 \ln\left ( 1 + e^{-\sqrt{u^2 + (m_i/T)^2}} \right )$ and $B$ is the bag
constant.
In Eq.~(\ref{pq}), the first term is due to massless gluons while
the second term needs to be evaluated for different physical masses
of the quarks. If the quarks are taken as massless, we get
\begin{eqnarray}
  p_q= a_q T^4 -B \hspace{.1in}{\rm where} \hspace{.1in}
    a_q= \frac{19\,\pi^2}{36}.
\end{eqnarray}
If the physical masses of quarks are used as
$m_u \simeq m_d \simeq 8$ MeV and $m_s \simeq 160$ MeV,
then at the critical temperature ($T_C=160$ MeV) we get
$f(m_u/T)\simeq 1$ and $f(m_s/T)\simeq 0.8$.
Thus,
\begin{eqnarray}
   a_q= \frac{37\,\pi^2}{90} + 0.8 \frac{7\,\pi^2}{60}.
\end{eqnarray}
which is not much different from the one for the massless gas.
Similarly, the
total pressure in hadron phase consisting of $n$ different types
of particles and resonances is given by
\begin{eqnarray}
p_h&=& - T \sum_{i=1}^n \, \frac{g_i}{2\pi^2} \int_{0}^{\infty}\,dk\, k^2
          \ln\left ( 1 - e^{-\sqrt{k^2 + m_i^2}/T} \right ), \nonumber \\
   &=& \frac{\pi^2}{90} T^4 \sum_{i=1}^n g_i h(m_i/T).
\end{eqnarray}
Here, $h(m_i/T)=- \frac{45}{\pi^4}
       \int_{0}^{\infty}\,du\, u^2
       \ln\left ( 1 - e^{-\sqrt{u^2 + (m_i/T)^2}} \right )$.
If we consider only a massless pion gas, the above summation contributes only
a factor of $3$. Assuming the hadron gas consisting of $\pi$,
$K$, $\eta$, $\rho$ and $\omega$ with appropriate mass and degeneracy
[$m_\pi$      = 137.0  MeV (g=3),
$m_K $        = 496.0  MeV (g=4),
$m_\eta$      = 547.45 MeV (g=1),
$m_\rho$      = 768.50 MeV (g=9),
$m_\omega$    = 781.94 MeV (g=3)]
we get
\begin{eqnarray}
p_h \simeq a_h  T^4, \hspace{.1in} {\rm where \hspace{.1in}}
  a_h = 4.5 \frac{\pi^2}{90} \hspace{.1in}
          {\rm at \hspace{.1in}}T = 160 {\rm MeV}.
\end{eqnarray}

In addition to the above equation of state, we use the following
phenomenological potential to describe the phase transition \cite{IGNAT}

\begin{eqnarray}\label{vt}
V(\phi, T) &=& a(T)\,\phi^2 - b \,T \,\phi^{3 } +  c \, \phi^4,
\end{eqnarray}
where $b$ and $c$ are  positive constants.
Although, the exact knowledge of the order parameter $\phi$ is
not necessary for the present purpose, it can be related to entropy
or energy density \cite{CSER1} or to the sigma and pion field \cite{DUMHEP}.
The above potential has two minima, one at $\phi_q=0$ and the other one at
\begin{equation} \label{phih}
\phi_h= (3 b T + \sqrt{9 b^2 T^2 -32 a c})/8c,
\end{equation}
which in the present case will represent quark and hadron phases respectively.
These phases are separated by a maximum that occurs at $\phi_m$ given by

\begin{equation} \label {phim}
\phi_m= (3 b T - \sqrt{9 b^2 T^2 -32 a c})/8c.
\end{equation}
In the thin wall approximation \cite{LINDE}, $b$ and $c$ can be
expressed in terms of surface tension $\sigma$ and correlation
length $\xi$ as \cite{INHOMO}

\begin {eqnarray} \label {con}
b=\frac{1}{\sqrt{6 \sigma \xi^5 T_C^2}};~c=\frac{1}{12 \xi^3 \sigma}.
\end{eqnarray}
The height of the degenerate barrier at $T=T_C$ or at
$a(T_C)=b^2T_C^2/4c$ is given by
\begin{equation}\label{bar}
V_b(T_C)=\frac{3\sigma}{16\xi (T_C)}.
\end{equation}
Thus, the potential can describe first-order phase transitions, with
varying strength by either changing $\sigma$ or $\xi$ or both.
The latent heat is also a measure of the
strength of the first order phase transition
which in the bag model is given by $4B$.
The requirement that at all the temperatures,
the difference between the two minima
should be equal to the pressure difference between
the two phases \cite{INHOMO} fixes the third parameter $a(T)$ as

\begin{eqnarray}\label{atemp}
\Delta p = p_h-p_q
         = -\left[V(\phi_h) - V(0)\right ] \nonumber \\
         =- \left[ a(T) - b T \phi_h + c \phi_h^2\right ]\phi_h^2.
\end{eqnarray}
Scavenius and Dumitru \cite{DUMHEP} have used a linear sigma model (LSM)
potential and fitted the potential of the form given by Eq.~(\ref{vt}) to
fix the parameters $a$, $b$ and $c$. With this form of potential they
obtain surface tension $\sigma$ and  correlation length $\xi$.
We directly use surface tension $(\sigma)$ and
correlation length $(\xi)$ to
fix the parameters $b$ and $c$. For any value of $b$ and $c$ the
nonperturbative vacuum effect is taken care of by bag constant $B$.
At $T=T_C$, $\phi_h=bT_C/2c=\sqrt{ 6 \sigma \xi}$. For $\sigma=10$ MeV/fm$^2$
and $\xi=0.7$ fm (typical hadron size)
we get $\phi_h=91$ MeV which is nothing but the pion decay constant.

Figure 1 shows the plot of $V(\phi)$ as a function of $\phi$ at four different
temperatures for typical values of
$\sigma$ = 30 MeV/fm$^2$ and $\xi(T_C) = 0.7$ fm.
At $T=T_C$, the potential is degenerate with $V(\phi_q)=V(\phi_h)=0$ being
separated by a barrier at $\phi=\phi_m$. As the matter supercools, the hadron
phase has lower free energy as compared to the QGP phase which is held fixed
at $\phi_q=0$ and also $V(\phi_q)=0$. Thus, below $T_C$, the $\phi_q$ phase is
metastable with respect to the stable hadron phase at $\phi=\phi_h$.
At $T=T_S$, the potential develops an inflection point where $\phi_q$ and
$\phi_m$ coincide. The condition $\phi_q=\phi_m=0$ leads to $a(T_S)=0$ and
$\phi_h=3b T_S/4c$ . The spinodal temperature $T_S$ can be obtained
analytically by solving Eq.~(\ref{atemp}) as
\begin{eqnarray}\label {tspin}
T_S = \left[\frac{B}{B + 27 V_b(T_C)} \right]^{1/4} \, T_C.
\end{eqnarray}
At $T=T_S$ their exists only one minimum corresponding to the hadron phase.
If the QGP supercools upto this point it will become unstable
and may go to hadron phase by spinodal decomposition.
It is worth noting that as $\sigma\rightarrow 0$ and
$\xi\rightarrow \infty$, $T_S\rightarrow T_C$.
The spinodal temperature depends on the strength of the
transition. For a strong enough transition, $\sigma$ is large
and $T_S$ is lower as compared to the case when the transition is weak.
We are interested to know whether the system cools down to the
temperature $T_S$. For comparison we denote the minimum temperature
reached during the supercooling as $T_m$. It is the temperature in the
supercooling region at which the system starts reheating due to the
release of latent heat.
 The rate of nucleation will be suppressed for a
stronger first order phase transition due to large barrier resulting
in higher supercooling, (i.e., smaller value of $T_m$).
Thus, both $T_S$ and $T_m$ depend on the strength of the transition and need
to be evaluated properly. While $T_S$ can be estimated directly
from Eq.~(\ref{tspin}), $T_m$ requires a self consistent solution of a
set of equations involving
the nucleation rate and energy momentum conserving hydrodynamical equations.

\begin{figure}
\centerline{\psfig{figure=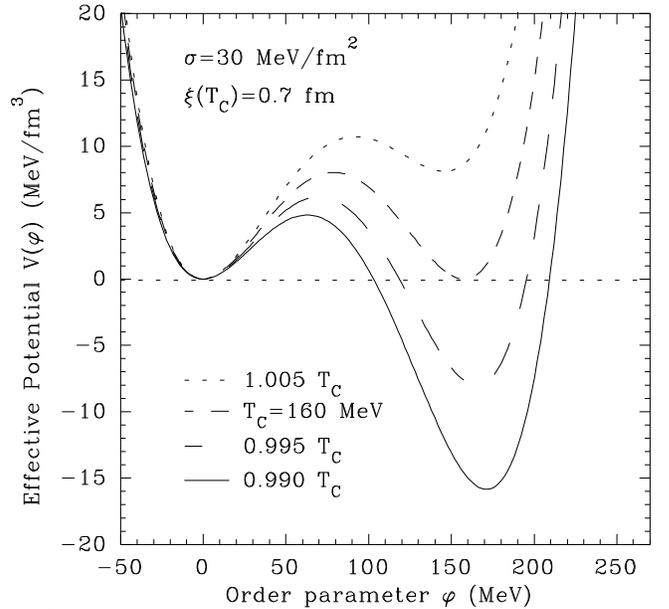,width=8.5cm,height=8cm}}
\caption{The free energy density  $V(\phi)$ as a function of $\phi$
at four different temperatures.
}
\end{figure}

\section{Nucleation rate}

The homogeneous nucleation theory assumes the formation of nucleating clusters
within the initially homogeneous metastable state. One can use the
Langer formalism \cite{LANG},
to calculate the probability of nucleation of hadron bubbles per
unit time per unit volume from a homogeneous background of quark gluon plasma,
given by
\begin{equation}
\displaystyle
I =  I_0 \, \exp{\left( - \frac{\Delta F_C}{T} \right)}\ .
\end{equation}
Here, $\Delta F_C$ is the minimum energy needed to form a  critical hadron
bubble and $I_0$ is the prefactor which can be written as
$I_0=\kappa  \Omega /2 \pi$.
The statistical factor $\Omega$
is a measure of both the available
phase space as the system goes over the saddle and of
the statistical fluctuations
at the saddle relative to the equilibrium states. The dynamical
prefactor $\kappa$ gives the exponential growth rate of the bubble or
droplet sitting on the saddle.

\subsection {Statistical prefactor}

To understand the meaning of the statistical
prefactor, consider a classical system with $N$ degrees of
freedom described by a set of $N$
collective coordinates $\eta_i,\ i = 1, \ldots , N$. The coarse-grained
free energy functional $F\{\eta\}$ of the system has local minima
$F\{\eta_i\}$ in the $\{\eta\}$-space, corresponding to the metastable
$\{\eta^0\}$ and stable states , separated by the energy barrier.
The point of minimal energy along the barrier is the so-called saddle
point $\{\eta^S\}$. The rate of the decay of the
metastable state is determined by the steady-state current across the
saddle point from the metastable to the stable minimum of $F\{\eta\}$.
According to \cite{LANG,CALLAN,AFF}, the statistical prefactor can be
written as
\begin{equation} \label{fluc}
\displaystyle
\Omega = \left( \frac{8 \pi \sigma}{3 |\lambda_1|}\right)^{3/2} \,
\left( \frac{2 \pi T}{|\lambda_1|} \right)
^{1/2}\, \left [ \frac{det (M_0/2\pi T)}{det(M \prime /2\pi T)}\right ] .
\end{equation}
where $\lambda_1$ is the negative eigen value of the matrix $M_{ij}=\partial^2F
/\partial \eta_i \partial \eta_j$, evaluated at the saddle point.
The subscript $0$ denotes the metastable state and the prime indicates
that the negative eigen value $\lambda_1$ as well as the zero eigen
values of the matrix $M_{ij}$ is omitted. The calculation of the fluctuation
determinant in Eq.~(\ref{fluc}) is extremely difficult and adds to the
uncertainty of the calculation of $\Omega$. In Ref.~\cite{LANGER73},
the above prefactor has been estimated under harmonic
approximation using the product of eigen values of the mobility matrix $M$,
evaluated at the saddle point and the metastable points.
The final expression that accounts for fluctuation correction reads

\begin{equation} \label{om1}
\displaystyle
\Omega_1 = \frac {2}{3 \sqrt{3}} \,
\left( \frac{\sigma}{T} \right)^{3/2} \,
\left( \frac{R_C}{\xi} \right)^{4} \ ,
\end{equation}
where $\sigma$ is the surface tension, $\xi$ is the correlation length
and $R_C$ is the size of the critical droplet. Since $R_C$ is infinite
at $T=T_C$, $\Omega_1$ diverges as $T\rightarrow T_C$. Alternatively, one
can follow the earlier approach of Langer \cite{LANG} where the
fluctuation corrections are absorbed into the free energies of the
metastable $(F_0)$
and saddle point region $(F_S)$ and the activation energy of the critical
droplet is simply $\Delta F_C=F_S-F_0$. Therefore, omitting the fluctuation
determinant, the expression for $\Omega$ is given by

\begin{equation}\label{om2}
\displaystyle
\Omega_2 = \frac{32 \pi^2 T^{1/2}}{|\lambda_1|^2}\,
\left( \frac{\sigma}{3} \right)^{3/2} \ .
\end{equation}
In the above, $\sigma$ is interpreted everywhere as the true surface
energy that includes fluctuation corrections. This approach has been
adopted in Refs.~\cite{ZABPL,ZABPRC} to estimate the statistical prefactor.
In order to evaluate $\lambda_1$, we will approximate the potential barrier
between the metastable and stable states by the excess of the free energy
$\Delta F$ corresponding to the formation of a spherical bubble of radius
$R$. In the thin wall approximation \cite{LINDE}, it can be written as
the sum of the bulk and the surface energies as

\begin{equation}\label{frt}
\Delta F(R,T) = -{4\pi \over 3} R^3 \Delta p + 4 \pi R^2 \sigma,
\end{equation}
where $\Delta p$ is the pressure difference between the hadron and QGP phases.
Minimization of $\Delta F$ with respect to the radius $R$ yields the free
energy of the critical bubble
\begin{equation}\label{crit}
\Delta F_C=\frac{4}{3} \pi R_C^2 \sigma;~~~ R_C=\frac{2\sigma}{\Delta p} .
\end{equation}
It is convenient to introduce a similarity number
$\lambda_z=R_C\sqrt{4\pi \sigma/T}$ and the reduced radius
$r=R/R_C$ \cite{ZABPL,ZABPRC,ZABPLA}. Two functions
$\Delta F_1(R)$ and $\Delta F_2(R)$ are similar if they have same
similarity number. In the harmonic approximation,
Eq.~(\ref{frt}) can be expanded around the critical radius as
\begin{eqnarray}
\frac{\Delta F}{T}& = &\left (\frac{\Delta F}{T} \right )_{R=R_C}
    +\frac{1}{2T} \left (\frac{\partial^2 \Delta F}{\partial R^2}
       \right )_{R=R_C} (R-R_C)^2\nonumber\\
 &=&\frac{\Delta F_C}{T} -\lambda_z^2(r-1)^2
\end{eqnarray}
and we get finally for the negative eigen value $\lambda_1$

\begin{equation} \label{lam1}
\lambda_1=-2T\lambda_z^2.
\end{equation}
Following \cite{ZABPL,ZABPRC}, we  estimate $\Omega_2$ using
Eq.~(\ref{om2}) and Eq.~(\ref{lam1}).

\subsection {Dynamical prefactor}

Csernai and Kapusta \cite{CSER1} have derived
a dynamical prefactor $\kappa$ by generalizing the coarse-grained effective
field theory of Langer to a relativistic case.
For a baryon-free plasma, where the thermal
conductivity vanishes, the dynamical prefactor is
found to depend on the viscosity coefficients and is given by

\begin{equation}\label{ck}
\displaystyle
\kappa_{CK} = \frac{4 \sigma \left(4\eta_q/3 + \zeta_q\right)}
{(\Delta \omega)^2\, R_C^3}\ .
\end{equation}
Here $\eta_q$ and $\zeta_q$ are the shear and bulk viscosity coefficients
of the quark phase
respectively and the $\Delta\omega$ is the difference between
the enthalpy densities of the QGP and the hadron phases,
$\omega=e+p$. According to Csernai and Kapusta (CK), there will be no bubble
growth in the case of an ideal plasma with zero viscosity.
Since viscosity is an essential ingredient in the above expression,
for consistency, dissipative hydrodynamics should also be used to
describe the space time evolution of the matter \cite{SHUK}.
The CK approach
has also been extended to include thermal dissipation in addition to viscous
damping for the case of baryon rich quark gluon plasma \cite{VEN}.
On the contrary, Ruggeri and Friedman \cite{RUGGERI} argued that for
relativistic hydrodynamics
the energy flow does not vanish in the absence of any heat conduction
or viscous damping. Since the change of the energy density $e$
in time is given in the low
velocity limit by the conservation equation \cite{CSER1},
$\partial e/\partial t = - \nabla . (\omega \bf v)$ which
implies that the energy flow $\propto \omega \bf v$ is always present.
Therefore, following a different approach,
Ruggeri and Friedman derived an expression
for $\kappa$, given by (for zero viscosity)

\begin{equation}\label{rug}
\displaystyle
\kappa_{RF} = \left( \frac{2 \sigma \omega_q}{R_C^3 (\Delta \omega)^2}
\right)^{1/2}.
\end{equation}
The above result is in contradiction with the expression given
by Eq.~(\ref{ck}). Recently, following the Langer's procedure, we have  solved
the relativistic hydrodynamic in the hadron, quark and the interfacial regions
to obtain the dynamical prefactor. The
$\kappa$ as obtained in Ref.~\cite{DYNA}, is given by

\begin{equation}\label{akm1}
\displaystyle
\kappa = \kappa_0 + \kappa_v; \\
 \kappa_0 = c_s \sqrt \frac{\xi}{3 R_C^3},
  \kappa_v =  \frac{\xi}{6 R_C^3} {1\over \omega_q}
  \left({4\eta_q\over 3} + \zeta_q \right).
\end{equation}
Here $c_s$ is the velocity of sound in the massless gas.
Under certain assumption for $c_s$, the above prefactor can also be
written as \cite{DYNA}

\begin{equation}\label{akm2}
\kappa \approx \left( \frac{2 \sigma \omega_q}{R_C^3 (\Delta \omega)^2}
\right)^{1/2}+ \frac{1}{c_s^2}
\frac{\sigma \left(4\eta_q/3 + \zeta_q\right)}
{(\Delta \omega)^2\, R_C^3} \ .
\end{equation}
 As can be seen, the first term in the above equation is the same
as obtained by Ruggeri and Friedman corresponding to the
case of nonviscous plasma [see Eq.~(\ref{rug})].
The second term is similar to the result
obtained by Csernai and Kapusta.
Two important aspects of our result are (a) the  prefactor $\kappa$ can be
written as a linear sum of a nonviscous ($\kappa_0$) and a
viscous ($\kappa_v$) components and (b) the nonviscous
component $(\kappa_0)$ which depends on two parameters $R$ and $\xi$ is
finite in the limit of zero viscosity. Further, it has been
argued in Ref.~\cite{DYNA} that the nonviscous part of the prefactor
basically arises  due to nonuniform pressure across the interface. In the
present work, we will use Eq.~(\ref{akm1}) to calculate the dynamical
prefactor.

\section{Nucleation and the expansion dynamics}

To calculate the fraction of space converted
to hadron phase using the nucleation rate $I(T)$, one
requires a kinetic equation.
The simplest of such equation is given in Ref. \cite{CSER2}.
According to it, if the QGP cools to $T_C$ at a proper time $\tau_C$,
then at some later time $\tau$ the fraction $h$ of space which has been
converted to hadronic gas is given by
\begin{eqnarray}\label{frac}
h(\tau) = \int_{\tau_C}^\tau d\tau' I(T(\tau')) [1 - h(\tau')] V(\tau',\tau).
\end{eqnarray}
Here $V(\tau',\tau)$ is the volume of a bubble at time $\tau$ which had
been nucleated at an earlier time $\tau'$; this takes into account
the bubble growth.
The factor $\left[ 1 - h(\tau')\right]$ is the available space for new
bubbles to nucleate.
The model for bubble growth can be taken as \cite{CSER2,WEIN}
\begin{eqnarray}\label{}
V(\tau',\tau) = \frac{4\pi}{3} \left( R(T(\tau')) +
\int_{\tau'}^\tau d\tau'' v(T(\tau'')) \right) ^3,
\end{eqnarray}
where $v(T)$ is the velocity of the bubble growth at temperature $T$
and is taken to be  $v(T) = 3 [1 - T/T_C]^{3/2}$ \cite{CSER2,MILLER}.
This expression is intended to apply only when
$T > \frac{2}{3} T_C$ so that the growth velocity stays below the speed
of sound of a massless gas, $c/\sqrt {3}$.
At the critical temperature, both the nucleation rate and growth rate
vanish. One can also write $v$ in terms of pressure
difference between the two phases as used in Ref. \cite{KAPVIS}.

The evolution of the energy density is given by  \cite{BJOR,HOSOYA,DAN}
\begin{eqnarray}\label{hydro}
\frac{d e}{d\tau} + \frac{D \,\omega}{\tau} = {4\eta/3 + \zeta \over \tau^2}
  \equiv {\mu \over \tau^2} \ ,
\end{eqnarray}
where $D$=1 for the expansion in $(1+1)$ dimension.
The factors $\eta$ and $\zeta$
are the shear and the bulk viscosity of the medium.
For nonviscous plasma (for zero viscosity), the above equation
follows the Bjorken's scaling solution where $T^3\tau$ is a constant.
The energy momentum
equation needs to be solved numerically for expansion in $(3+1)$
dimensions. However, retaining the simplicity,
we can still use Eq.~(\ref{hydro})
for spherical expansion \cite{KAPZ,COOPER}
with the choice of $D=3$,
although the
scaling $u^\mu=x^\mu/\tau$ (where $u^\mu$ is the four velocity of
the fluid) may not be  valid
if viscosity is included.

  In the transition region, the energy density at a time $\tau$
can be written in terms of hadronic fraction $h(\tau)$ as
\begin{eqnarray}\label{}
e(\tau) & = &  e_q(T) + [e_h(T)-e_q(T)]
       h(\tau).
\end{eqnarray}
Here $e_q= 3 p_q + 4 B$  and $e_h= 3 p_h$ are energy densities in the
QGP and hadron phase respectively. The viscosity at any time $\tau$
can also be written \cite{JPG} as
\begin{eqnarray}\label{}
\mu(\tau) & = &  \mu_q(T) + [\mu_h(T)-\mu_q(T)] h(\tau),
\end{eqnarray}
where $\mu_q=4\eta_q/3 + \zeta_q$ and $\mu_h=4\eta_h/3 + \zeta_h$.
The temperature dependence of the viscosities are taken as
$\sim T^3$ \cite{HOSOYA}.

\section{Results and Discussions}
We compare the nucleation rates using
different dynamical prefactors $\kappa$ as given by

\begin{eqnarray}
I_1 &=& \frac{\Omega_2 \kappa_{CK}}{2\pi} \,
       \exp{\left( - \frac{\Delta F_C}{T} \right)}\ ,\nonumber\\
I_2 &=& \frac{\Omega_2 \kappa_0}{2\pi} \,
       \exp{\left( - \frac{\Delta F_C}{T} \right)}\ ,\nonumber\\
I_3 &=& \frac{\Omega_2 \kappa_v}{2\pi} \,
      \exp{\left( - \frac{\Delta F_C}{T} \right)}\ ,\nonumber\\
I_4 &=& \frac{\Omega_2 (\kappa_0+\kappa_v)}{2\pi} \,
       \exp{\left( - \frac{\Delta F_C}{T} \right)}\ ,\nonumber\\
I_5 &=& T^4 \, \exp{\left( - \frac{\Delta F_C}{T} \right)}\ .
\end{eqnarray}

In all the cases, we use the same statistical prefactor $\Omega_2$
which is more physical
except for $I_5$ where the total prefactor is of the order of $\sim T^4$
as commonly used in the literature \cite{LINDE}. Further,
we choose $\xi=0.7$ fm and use Eq.~(\ref{crit}) to
estimate $\Delta F_C/T$. The nucleation
rates $I_2$ and $I_5$  correspond to the case of a nondissipative plasma
whereas viscosity enters as an
essential ingredient for the evaluation of $I_1$, $I_3$ and $I_4$.
Figures 2 and 3 show the plot of
above nucleation rates as a function of $T/T_C$ for $\sigma=10$ MeV/fm$^2$ and
$\sigma=30$ MeV/fm$^2$ respectively each
at two typical values of viscosity coefficient $\eta_q$.
We consider only the shear viscosity $\eta_q$ since the bulk
viscosity $\zeta_q$ that provides resistance to expansion (or contraction)
does not exist in the case of an incompressible fluid. Further,
the viscosity coefficients $\eta_h$ and $\zeta_h$ in hadron phase are
neglected since the amount of supercooling is affected by viscosity
of quark phase only. Notice that the
most commonly used prefactor $\sim T^4$ overpredicts the
nucleation rate ($I_5$) over a wide range of temperature particularly
for small amount of supercooling and
also when the transition is relatively stronger (see Fig.~3).
Further notice, that both for strong and weak
transitions but for moderately viscous medium, the nucleation rates are
dominated by the nonviscous components (since $I_4$ is nearly the same as
$I_2$ , see Figs~2(a) and 3(a) for $\eta_q=5 T^3$).
The rates $I_1$ and $I_3$ which contain viscosity have only a second order
effect unless the medium is highly dissipative. At $\eta_q=14.4 T^3$,
the viscous component $I_3$ competes
with the nonviscous component $I_2$ although $I_1$ is still lower.
[see Figs.~2(b) and 3(b)].
From the above studies we may conclude that the choice of $T^4$ overestimates
whereas the rate $(I_1)$ with the Csernai-Kapusta prefactor underestimates
the nucleation rates in the region of interest as compared to $I_4$. Since
the nucleation rate $I_4$
has both dissipative and nondissipative components, we will use
it to estimate the nucleation rate and supercooling in the subsequent study.

\begin{figure}
\centerline{\psfig{figure=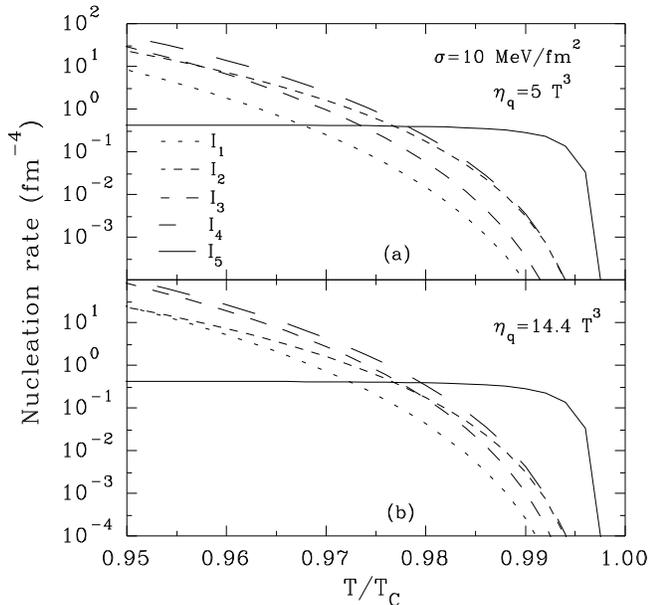,width=8.5cm,height=8cm}}
\caption{Nucleation rates obtained using different prefactors
for two viscosities using surface tension $\sigma=10$ MeV/fm$^2$.
The expressions for $I_1$ etc. are given in the text.
}
\end{figure}

\begin{figure}
\centerline{\psfig{figure=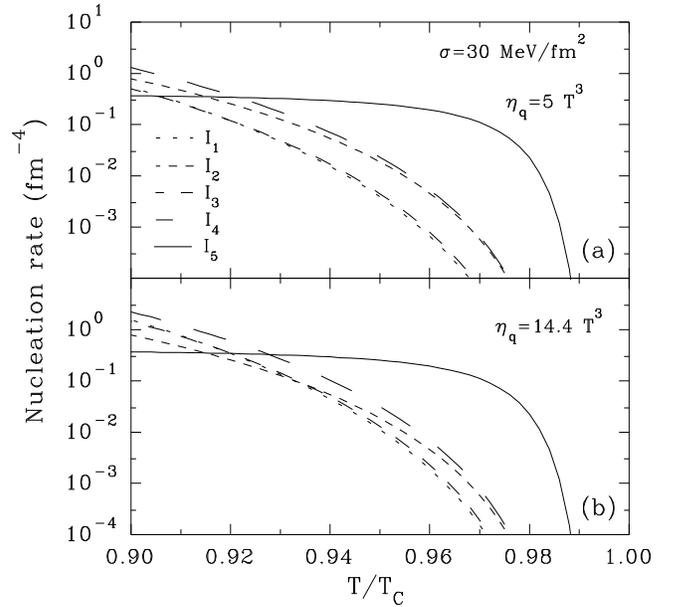,width=8.5cm,height=8cm}}
\caption{Same as Fig.~2 but with $\sigma=30$ MeV/fm$^2$.
}
\end{figure}

Next, we solve the coupled equations Eq.~(\ref{frac}) and
Eq.~(\ref{hydro}) to study the space time evolution of the matter.
The critical temperature is fixed at $T_C$ = 160 MeV
which gives $B^{1/4}=236$ MeV.
The strength of the transition depends on $\sigma$ and $\xi$ or
more explicitly on the ratio $\sigma/\xi$. Recent lattice
calculations \cite{IWA94} predict $\sigma$ between 2 to 10 MeV/fm$^2$.
These calculations are without any dynamical quarks and thus are only
indicative. Therefore $\sigma$ is treated  as a free
parameter in the present study and $\xi=0.7$ fm seems the
most reasonable value for the correlation length.

Another important parameter that affects the
evolution is the time $\tau_C$ that the plasma takes to cool down to $T=T_C$.
Obviously, $\tau_C$ will depend on the initial temperature $(T_i)$, the
formation time $(\tau_i)$ as well as on the expansion dynamics of the plasma.
In the Bjorken scenario \cite{BJOR}, $\tau_C=(T_i/T_C)^3 \tau_i$.
It is quite reasonable to assume that the plasma will expand
in $(1+1)$ dimension between $\tau_i$ and $\tau_C$ until the
longitudinal dimension becomes comparable to the size of the colliding nuclei.
Assuming $T_i=320$ MeV and $\tau_i = 1$ fm/$c$, the Bjorken scaling predicts
$\tau_C=8$ fm/$c$  for expansion in $(1+1)$ dimension.
%It is also expected that
%the plasma produced at RHIC and LHC energies will attain a very high initial
%temperature in a short time. Such a plasma if formed
%will have high gluon contents \cite{SHUR} and many more quark and anti-quark
%will be needed in order to complete chemical equilibrium. A chemically
%equilibrating plasma will cool still at a faster rate as compared to the
%normal plasma \cite{BIRO,DUTTA}.
As mentioned before, $\tau_C$ is a
crucial parameter that affect the solution of the coupled equations which
in turn depends on the initial conditions as well as on the
expansion scenario of the plasma. Since $\tau_C$ is not known exactly,
we assume it of the order of  6 to 8 fm/$c$.
From $\tau_C$ onwards, we consider the spherical
expansion. This expansion scenario corresponds
to the fastest cooling rate resulting in maximum supercooling. Therefore the
choice of $D=3$ provides a lower bound on $T_m$ which we compare with
the spinodal temperature $T_S$.

 Figure 4 shows the plot of $T/T_C$ as a
function of $\tau$ for a nonviscous plasma at a typical value of
$\tau_C=8$ fm/$c$. Unlike the ideal Maxwell construction ( where the
temperature is held fixed at $T=T_C$ until phase conversion is over), the
system supercools upto $T_m$. At $T_m$, the number density of the
nucleated hadron bubbles is sufficient to raise the temperature again
due to the release of the latent heat. The amount of supercooling
will obviously depend on the strength of the transition being more for a
stronger transition (large value of $\sigma$) as compared to that
for the weaker one
(small value of $\sigma$). A faster expansion also reduces the number density
of the nucleated hadron bubbles resulting in a larger supercooling. Although
shown for $\tau_C=8$ fm/$c$, the amount of supercooling will increase
further if  $\tau_C$ is reduced due to faster expansion. We have also
done the calculations for $\tau_C=6$ fm/$c$ which are summarized in
Fig.~6. For $\tau_C$ still smaller the expansion in (1+1) should be
used.

\begin{figure}
\centerline{\psfig{figure=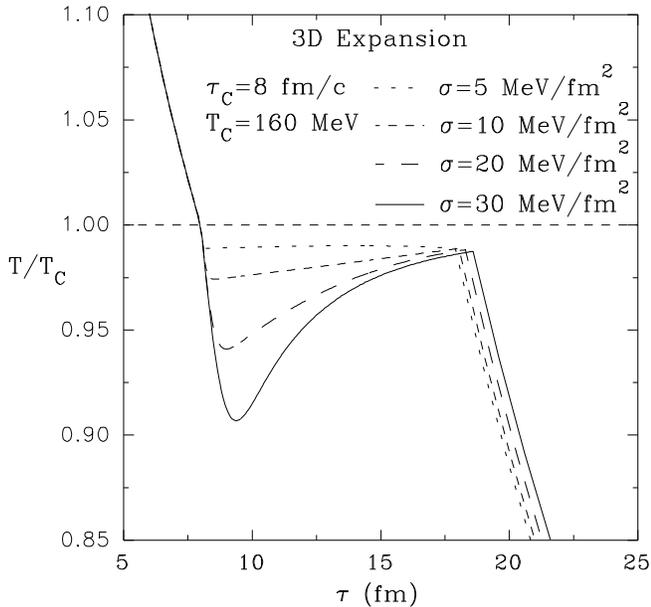,width=8.5cm,height=8cm}}
\caption{The temperature evolution obtained with the
nucleation rate $I_{4}$ for a spherically expanding system
for different surface tensions.
}
\end{figure}

 Next, we consider the dissipative hydrodynamics. As mentioned before, the
viscous contribution to the nucleation rate is not very significant if the
transition is strong enough and also the medium is moderately viscous. However,
the presence of a small amount of viscosity will affect the hydrodynamic
evolution of the plasma. The temperature of the plasma will fall at a slower
rate due to viscous heating of the medium. Figure~5 shows a plot
of $T/T_C$ as a function of $\tau$ for $\eta_q$=$0$, $5T^3$ and $14.4T^3$.
Here we have considered the case of a relatively
stronger transition ($\sigma=30$ MeV/fm$^2$) where the viscosity
contribution to the nucleation rate is not very significant. However,
significant effect can be seen on $T_m$ which increases (less supercooling)
with increasing
viscosity as expected \cite{SHUK}. 

\begin{figure}
\centerline{\psfig{figure=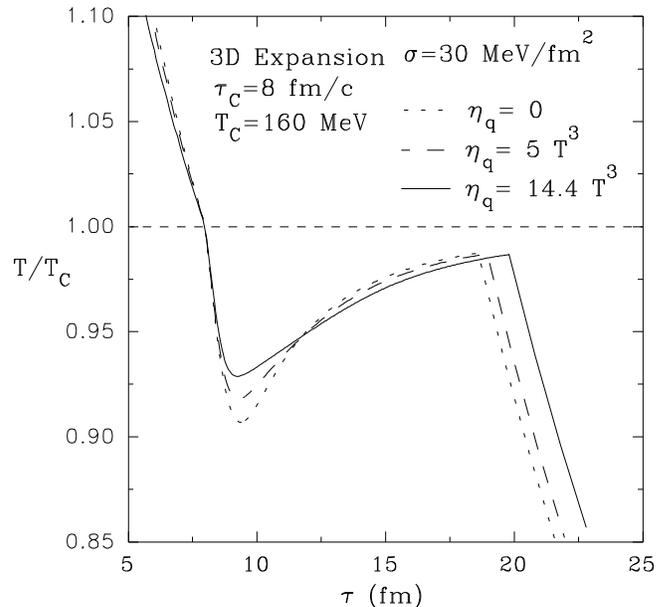,width=8.5cm,height=8cm}}
\caption{The temperature evolution obtained with the
nucleation rate $I_{4}$ for a spherically expanding system
for different viscosity coeffs.
}
\end{figure}

Figures~6 and 7 show the
plot of $T_m/T_C$ as a function of $\sigma/\xi_q$ at $\tau_C$=
$6$ fm/$c$ and $8$ fm/$c$ respectively both for ideal as well as
for dissipative plasma.
The amount of supercooling becomes less as the medium becomes more viscous.
The spinodal temperature $T_S$ which depends on the ratio
$\sigma/\xi$ has also been plotted in the same figures for comparison.
First consider the case of
nondissipative plasma with zero viscosity. The curves $T_m$ and $T_S$ show
a cross over point depending on the choice of $\tau_C$. The nucleation rate
is suppressed for a strong enough transition due to large nucleation barrier
resulting in a higher amount of supercooling. On the other hand for
weak enough transition, the amount of supercooling is smaller and also
well above the point of spinodal decomposition. Since $T_m$ depends on
expansion rate of the medium, the cross over point will sensitively
depend on $\tau_C$; moving towards left for the faster expansion.
Qualitatively it can be concluded here that the homogeneous
nucleation is still the dominant mechanism of phase conversion
if the transition is weak enough. For stronger transition and fast
enough cooling, the system may reach $T_S$ and the phase
conversion may proceed through the spinodal decomposition.
If the plasma is viscous, the amount
of supercooling is reduced further due to slow evolution of the medium.
Even the cross over point also shifts towards
right showing that the nucleation is the dominant mechanism over a wide range
of $\sigma/\xi$ ratios.
Due to uncertainties in $\tau_C$, $\sigma/\xi$ and $\eta_q$ ,
it is difficult to say what is the exact scenario at RHIC and LHC
energies.
Assuming a weakly first order transition for $\sigma$ in the range
of 2 to 5 MeV/fm$^2$, $\xi \sim 0.7$ fm \cite{IWA94}, $\eta_q \sim 2T^3$
(lower limit \cite{note}) and $\tau_C$ $\sim$ 4 to 8 fm/$c$ \cite{DUTTA},
the nucleation still seems to be the dominant mechanism of phase  conversion
as opposed to the spinodal decomposition.

It may be mentioned here that we have considered a dynamical growth rate
which is exponential in nature. This may not be true at the later stage of
the expansion. The fusion of bubbles may also play a role at the later stage
when the density of bubbles is sufficiently
high \cite{KAPZ}. However, in the study of supercooling, we
are very much in the domain of exponential region
where the effect of bubble fusion can be ignored.

\begin{figure}
\centerline{\psfig{figure=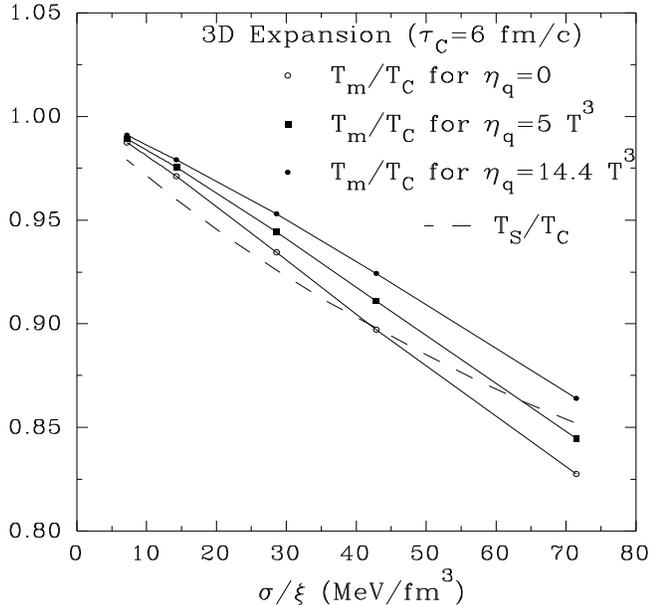,width=8.5cm,height=8cm}}
\caption{The spinodal temperature $T_S/T_C$ and minimum
temperature $T_m/T_C$ reached during supercooling as
a function of $\sigma/\xi$ using $\tau_C=6$  fm/c.
}
\end{figure}

\begin{figure}
\centerline{\psfig{figure=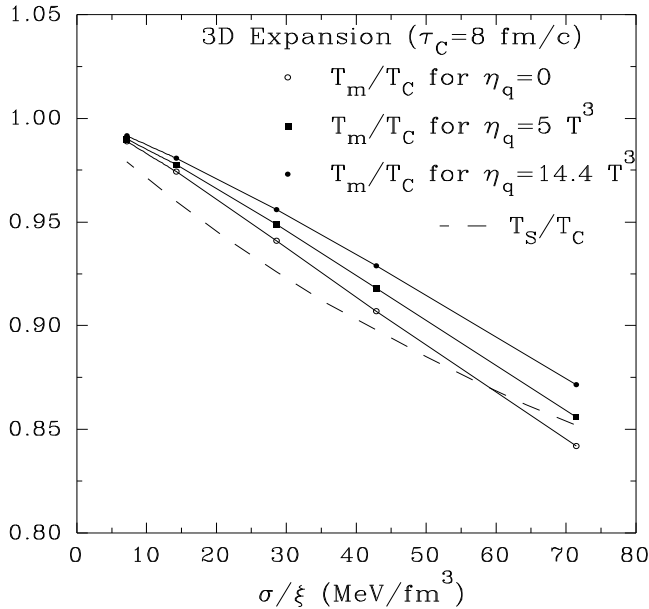,width=8.5cm,height=8cm}}
\caption{Same as Fig.~6 but with $\tau_C=8$  fm/c.
}
\end{figure}

\section {Conclusions}
We have investigated the mechanism of phase conversion from quark gluon
plasma phase to hadron phase via two routes: the standard homogeneous
nucleation and spinodal decomposition. The point of spinodal decomposition
depends on the strength of the transition or more precisely on the ratio
$\sigma/\xi$ where $\sigma$ and $\xi$ are the surface tension and correlation
length respectively at $T=T_C$. The nucleation and supercooling, on the
other hand, depend on the strength of the transition as well as on the
expansion dynamics of the medium. We have solved a set of coupled
equations to estimate the amount of maximum supercooling under spherical
expansion scenario. Which way the hadronization
will proceed depends sensitively on the nucleation and expansion dynamics
since the nucleation and expansion time scales are comparable.
 Qualitatively we can describe the results as follows:
For strong enough transition with zero or very small amount
of viscosity, the system reaches the spinodal instability before
the amount of nucleated hadron bubbles become significant to begin phase
conversion. The phase conversion in such a case
will proceed via spinodal decomposition.
If the medium is viscous or the transition is weak enough or both,
the supercooling is much less; The phase conversion may still proceed through
homogeneous nucleation. However, depending on the range of the parameters,
there could be a competition between the homogeneous nucleation and the
spinodal decomposition. A definite answer
can be provided only when the parameters
such as surface tension, $\tau_C$, viscosity and also the expansion
scenario are known precisely.


\begin{thebibliography}{99}
\bibitem{note}The acceptable range of $\eta$ for the application
              of scaling Navier-Stokes theory to the expansion
              of the plasma is given by $2 T^3 \le \eta \le 3 T^3 (\tau T)$.
              For detail see Ref.~\cite{DAN}
\end{thebibliography}
\end{document}